# Towards the statistical construction of hybrid development methods

Paolo Tell[1(✉)], Jil Klünder[2], Steffen Küpper[3], David Raffo[4], Stephen MacDonell[5], Jürgen Münch[6], Dietmar Pfahl[7], Oliver Linssen[8], and Marco Kuhrmann[9]

[1]*Department of Computer Science, IT University, Copenhagen, Denmark*
[2]*Software Engineering Group, Leibniz University Hannover, Hannover, Germany*
[3]*Center for Digital Technologies, Clausthal University of Technology, Clausthal-Zellerfeld, Germany*
[4]*Department of Computer Science, School of Business, Portland State University, Portland, Oregon, USA*
[5]*Department of Information Technology and Software Engineering, Auckland University of Technology, Auckland, New Zealand*
[6]*Department of Computer, Science, Reutlingen University, Reutlingen, Germany*
[7]*Institute of Computer Science, University of Tartu, Tartu, Estonia*
[8]*FOM Hochschule für Oekonomie & Management, Essen, Germany*
[9]*Software Engineering I, University of Passau, Passau, Germany*

**Abstract**

*Hardly any software development process is used as prescribed by authors or standards. Regardless of company size or industry sector, a majority of project teams and companies use hybrid development methods (short: hybrid methods) that combine different development methods and practices. Even though such hybrid methods are highly individualized, a common understanding of how to systematically construct synergetic practices is missing. In this article, we make a first step towards a statistical construction procedure for hybrid methods. Grounded in 1467 data points from a large-scale practitioner survey, we study the question: What are hybrid methods made of and how can they be systematically constructed? Our findings show that only eight methods and few practices build the core of modern software development. Using an 85% agreement level in the participants' selections, we provide examples illustrating how hybrid methods can be characterized by the practices they are made of. Furthermore, using this characterization, we develop an initial construction procedure, which allows for defining a method frame and enriching it incrementally to devise a hybrid method using ranked sets of practice.*

# 1. INTRODUCTION

Today, companies often use highly individualized processes to run projects, often by integrating agile methods in their processes. For instance, Dikert et al[1] found choosing and customizing an agile model to be an important success factor and that agility in general changed the way software is developed. Dingsøyr et al[2] reflect on a decade of agile methodologies, and there is no denial that agile methods have become an important asset in many companies' process portfolios.[3-6] However, agile methods are not implemented as prescribed by authors or standards,[7,8] and in 2011, West et al[9] coined the term "Water-Scrum-Fall" to describe a pattern that they claimed most companies implement for their software projects.

In previous research,[10,11] we could confirm West's claim. In addition, independently conducted research[12] and a number of country-specific[4,13] and industry-hosted studies[14] provide evidence on the use of *hybrid development methods*[10,15] (short: hybrid methods). Although agile methods have been an important research topic, they have also stimulated increasing diversity in software and system development, little information is available about the nature of hybrid development methods, what they look like, and how to devise them.

## 1.1 Problem statement

Modern software and system development does not follow any blueprint. A variety of different frameworks, methods, and practices are used in practice; according to a study by Klünder et al,[16] 78.5% of practitioners evolve their processes over time to improve, for instance, different product quality attributes and to keep flexibility regarding the ability to react to change. However, an understanding of what a hybrid development method is composed of is missing, for example, which combinations of frameworks, methods, and practices for software and system development help practitioners implement a process environment that provides the company and the management with a stable framework while providing developers with the demanded flexibility.[6,11]

## 1.2 Objective

The work presented in this paper aims to lay the foundation for understanding *hybrid development methods* and to develop adaptable construction procedures that help devise such methods grounded in evidence. The objective of our research is to *understand which frameworks, methods, and practices are used to realize hybrid methods in practice and to provide an evidence-based characterization of such methods.*

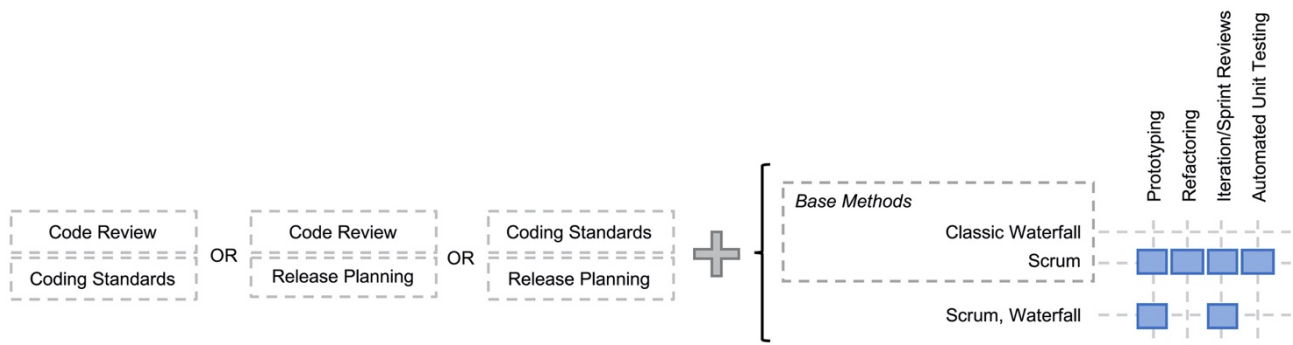

**Figure 1.** An example of a constructed hybrid method as shown in Tell et al.[18] that illustrates West's *Water-Scrum-Fall*.[9] The figure shows the base methods, the statistically computed core practices and the (*Scrum-Waterfall*) method, complemented with the different practices that shape the hybrid method, for example, *Prototyping*

### 1.3 Contribution

Based on a large-scale international online survey, we analyze 1467 data points that provide information about the combined use of 60 frameworks, methods, and practices. Our findings indicate that using hybrid development methods *is* the norm in modern software and system development and that using hybrid methods happens in companies of all sizes and across all industry sectors. We identify eight base methods providing the basis for devising hybrid methods, and we statistically compute sets of practices used to embody the base methods. We contribute a statistical process that helps computing hybrid methods (including process variants) to provide advice to practitioners what (not) to include in their process portfolio.

### 1.4 Context

The research presented in this paper emerges from the HELENA (Hybrid dEveLopmENt Approaches in software systems development, online: https://helenastudy.wordpress.com) study, which is a large-scale international online survey in which a team of 75 researchers and practitioners from 25 countries collected data world wide. The study was implemented in two stages (Figure 2) of which the first stage was a public trail executed in 2016 in Europe.[10, 15] All data and complementing materials of the second stage are available online.[17]

This article is an extended version of the previously published ICSSP conference paper[18] in which we studied the statistical construction of hybrid methods. For instance, Figure 1 shows how such a construction works: hybrid methods are computed based on statistically constructed base methods, method combinations, and core practices according to level of agreement within the dataset. These hybrid methods include a method combination as framework to host different practices. Due to the availability of different core practice clusters and several extra practices reported relevant by the HELENA study participants, in our previous paper,[18] we constructed hybrid methods and variants of such methods. The extension of our previously published paper is thus shaped by (1) an improved discussion of our work in the context related contributions and, moreover, (2) an extended analysis of the data. Specifically, we extended our discussion on how to utilize the analysis procedure documented in the conference paper[18] and derived an initial proposal for a *statistical construction procedure for hybrid development methods*. A new research question covers the necessary extra analyses, which aim to better characterize combinations of frameworks, methods, and practices forming hybrid development methods.

### 1.5 Outline

The paper is organized as follows: Section 2 presents related work. In Section 3, we present the research design. The results are presented in Section 4 and discussed in Section 5. The paper is concluded in Section 6.

## 2. RELATED WORK

The use of software development processes has been studied since the 1970s, when the first ideas to structure software development appeared.[19, 20] Since then, a growing number of approaches emerged, ranging from traditional and rather sequential models, to iterative and agile models

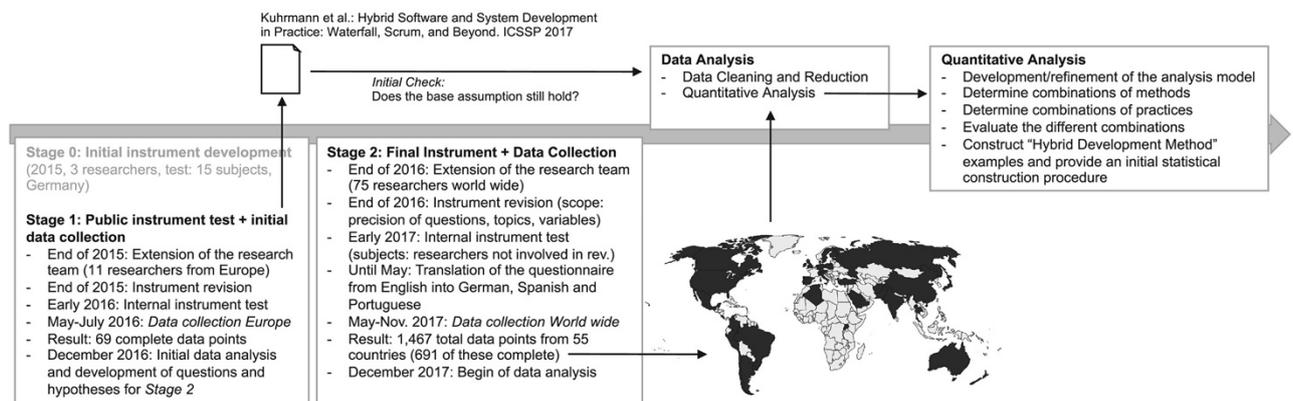

**Figure 2.** Overview of the research method applied including this study's position in the overall HELENA project and its previously published outcomes. The figure also includes a world map that highlights all 55 countries that contributed to the HELENA dataset

and a number of survey studies have sought to investigate the state of practice by focusing on software development methods. For instance, the "State of Agile" survey[14] annually collects data on the use of agile methods. The "Swiss Agile Study"[13] and the "Status Quo Agile" study[21] collect data in certain intervals aiming at observing the use of agile methods in Switzerland and Germany. Garousi et al[4] provide an overview of the use of agile methods in Turkey. Tripp and Armstrong[22] investigated the "most popular" agile methods and found XP, Scrum, Dynamic Systems Development Method (DSDM), Crystal, Feature Driven Development (FDD), and Lean Software Development among the top methods used. These studies, however, explicitly focus on agile methods and cover only some of the traditional methods. Dingsøyr et al[2] provided an overview of "a decade of agile software development" and motivate research towards a rigorous theoretical framework of agile software development, specifically, on methods of relevance for industry. However, Theocharis et al[11] provided evidence that this focus on agile is too narrow as, for instance, numerous companies and project teams remain skeptical and do not consider agile methods as the "Silver Bullet."[6,23-25] For example, Cusumano et al[26] surveyed 104 projects and found many using and combining different development approaches. In an analysis of 12 000 projects, Jones[27] found that both specific design methods and programming languages can lead to successful or troubled project outcome. Neill and Laplante[28] found that approximately 35% of developers used the classical Waterfall model. However, projects also used incremental approaches, even within particular lifecycle phases.

It is a matter of fact that companies develop a heterogeneous process portfolio comprised of a variety of traditional and agile methods and practices. For instance, Cockburn[29] describes a framework to choose appropriate methods to address the needs of projects. Boehm and Turner[30] aimed to overcome situation-specific shortcomings of agile and plan-driven development by defining five factors that describe a project environment and help determine a balanced method. Different complementary research streams were developed to address the required process variability and adaptability demands. For instance, Clarke and O'Connor[31] and Kalus and Kuhrmann[32] provide collections of situational factors or tailoring criteria to support process tailoring. Another research stream is focused on software process lines,[33,34] which aim to provide a comprehensive framework that allows for defining composite development methods, which comply with a standard/reference process model. Further initiatives to provide a structured approach towards modeling simple and complex heterogenous development methods are, among other things, reflected in initiatives such as SEMAT and Essence,[35-37] the standard ISO/IEC 24744:2007,[38] and numerous activities in the context of the SPEM software process engineering metamodel.[39,40] All these initiatives aim to bring more flexibility to processes and to help companies devise context-specific processes. Solinski and Petersen[41] aimed to characterize such combined processes and found Scrum and XP to be the most commonly adopted methods, with Waterfall/XP and Scrum/XP as the most common combinations. For such combined processes, West et al[9] coined the term "Water-Scrum-Fall", and different studies[11,12,42] provide evidence that the use of hybrid methods has become the norm. In 2017, we generalized this concept, defining the term "hybrid development methods" as "any combination of agile and traditional (plan-driven or rich) approaches that an organizational unit adopts and customizes to its own context needs."[10]

Available studies thus show a situation in which traditional and agile approaches coexist and form the majority of practically used hybrid methods. In contrast, current literature on software processes and their application in practice leaves researchers and practitioners with an increasing amount of research focusing on agile methods only. Traditional models are vanishing from researchers' focus. They only play a role in process modeling, in domains with special requirements (e.g., regulations and norms), or in discussions regarding the reasons why certain companies do not use agile methods.[6,22] In our previously published papers,[10,15] we initially studied the state of practice in using different frameworks, methods, and practices in combination and derived process clusters that form hybrid development methods. Klünder et al[16] studied the development of hybrid methods and found an evolutionary approach to be the common way to devise such methods, followed by planning a method as part of software process improvement programs.

The article at hand extends the available research by investigating the characteristics of hybrid development methods with a specific focus on the components of hybrid methods. We analyze combinations of frameworks, methods, and practices statistically to find such combinations that have a high level of agreement among the study participants and, thus, can be considered common sense about the basic structure of hybrid methods. Beyond the plain analysis, we also make a first step towards constructing hybrid methods by describing a statistical procedure that helps in computing hybrid methods and their variants from data.

## 3. RESEARCH DESIGN

We present the research design including the research questions, information about the survey instrument, and the different procedures regarding data collection, analysis, and the validity. The overall research design is outlined in Figure 2; the individual steps are described in the following paragraphs. Figure 2 also highlights the relation to the initial publication,[10] which was published based on the results of the first stage of HELENA and that lays the foundation for our original ICSSP conference paper[18] and the article at hand.

### 3.1 Research objective and research questions
The overall objective of the research presented in this paper is to understand which frameworks, methods, and practices are used to realize hybrid development methods in practice, to provide an evidence-based characterization of such methods, and to develop a *construction procedure*, which is grounded in evidence and allows for systematically constructing hybrid development methods. For this, we study the following research questions:

**RQ1:** *Which frameworks and methods form the basis for devising hybrid development methods?* This question sets the scene by analyzing the more comprehensive frameworks and methods that form the basis for hybrid methods and bind the different (smaller) practices together. This research question is motivated by a finding from our previous study10 that process clusters are formed around "centers." The first step is thus to identify such centers. As the HELENA study contains a flag that indicates if a specific set of frameworks, methods, and practices is intentionally used in combination, the analysis is performed twice: once for the entire dataset and once for the subset of data for which the study participants explicitly stated to combine the different processes.

**RQ2:** *Which practices are used to embody method combinations for devising hybrid development methods?* Having identified the base methods and the method combinations providing the frame for a hybrid method, we analyze the data for recurring practices used to embody the identified base methods and method combinations. That is, we aim to identify specific combinations of frameworks, methods, and practices that, together, form hybrid development methods. Again, the investigation is performed twice for the entire dataset and the subset of partici- pants that explicitly combine processes.

**RQ3:** *Which process variants are more promising?* The results from Tell et al.18 showcased a significant variability in the combinations of practices within each hybrid method leading to additional questions. For instance, is it possible to identify preferred process variances within the dataset? Therefore, we aim to further leverage the construction process18 to qualify process variability and to improve our understanding of promising process variants, that is, practitioners' preferred framework, method, and practice combinations.

**RQ4:** *How can hybrid development methods be characterized?* In Tell et al.,[18] we drafted an initial procedure with which we utilize the analysis procedures applied to the HELENA dataset to statistically construct hybrid methods. Including the findings from RQ 3, this research question aims to improve that procedure to help characterize hybrid methods by defining core practices that, together with the base methods and method combinations, provide a means to devise hybrid methods. Hence, we aim to statistically define hybrid methods, to define a hybrid method and its variant space to provide a ranked list of suitable hybrid methods and, eventually, to help practitioners decide what to (not) include into their process portfolio.

### 3.2 Instrument development and data collection

Data were collected using the survey method.[43] We designed an online questionnaire to solicit data from practitioners about the development approaches they use in their projects. The *unit of analysis* was either a project (ongoing or finished) or a software product.

#### 3.2.1 Instrument development and structure

The survey instrument was developed and refined in several iterations, which are illustrated in Figure 2. Finally, the research team included 75 researchers from all over the world. The questionnaire was made available in English, German, Spanish, and Portuguese and consisted of five parts (with number of questions[1]): Demographics (10), Process Use (13), Process Use and Standards (5), Experiences (2), and Closing (8). In total, the questionnaire composed up to 38 questions, depending on previously given answers. A complete overview of the questionnaire structure, the questions, variables and datatypes, and the conditional paths through the questionnaire is available from the accompanying research kit,[17] which also includes the raw questionnaire designs and their translations into the different target languages for independent replications and the final raw dataset.

#### 3.2.2 Data collection

Data were collected from May to November 2017 following a *convenience sampling strategy*.[43] The survey was promoted through personal contacts of the participating researchers, posters at conferences, as well as posts to mailing lists, social media channels (Twitter, Xing, LinkedIn), professional networks, and websites (ResearchGate and researchers' individual and/or institution home pages).

### 3.3 Data analysis procedures

As outlined in Figure 2, the data analysis consisted of multiple parts, which are described in detail in this section.

#### 3.3.1 Data cleaning and data reduction

The first step was the preparation of the data. We opted for the full dataset of the second stage of the HELENA study,[17] which consists of 1467 data points. As many questions were optional and participants had the opportunity to skip mandatory questions, we first analyzed the data for *NA* (i.e., "not available") and −9 values. While *NA* values indicate that participants did not provide information for an optional question, −9 indicates that participants skipped a mandatory question. Depending on the actual question, −9 values were either transformed into *NA* values or the respective data point was excluded from further analysis because we considered the question not completely answered. Finally, in the question about company size (question D001[17]), we combined the categories *Micro* and *Small* into a new category *Micro and Small (1–50 employees)* that resulted in an almost even distribution among all company sizes.

#### 3.3.2 Checking the base assumptions

In this study, we are interested in the particular process combinations used in industry. Our base assumption is that frameworks, methods, and practices are combined in practice as claimed by West et al[9] For this, in our previous studies,[10,15] we quantitatively analyzed the initial data using a set of hypotheses. As the first step in the quantitative data analysis, we tested the two hypotheses shown in Table 1 using Pearson's $\chi^2$ test at a significance level of 0.05.

---

[1] An important aspect to note is that no definition was provided for any given item, be this a method, a practice, or a technical concept in general. This was done to avoid any bias that might have been introduced from exposing participants to our knowledge, informing them on the concepts of interest, or leading them towards specific attitudes when answering the survey questions.

**Table 1.** Null hypotheses used to check the base assumption that combinations are common practice

| Hypotheses | | Question/variable assignment to hypotheses |
|---|---|---|
| $H1_0$ | The use of hybrid methods does not depend on the company size. | Combination (PU04), Company size (D001) |
| $H2_0$ | The use of hybrid methods does not depend on the industry target domain. | Combination (PU04), Industry sector (D005) |

While H1 was directly tested using Pearson's $\chi^2$ test, testing H2 required a different procedure as participants were able to provide more than one industry sector as targets for the question D005.[17] Therefore, a Pearson's $\chi^2$ test was evaluated for all industry sectors. For each industry sector, we tested the share of participants stating that they (do not) combine the different frameworks, methods, and practices and compared those with all the other industry sectors. As the number of data points per industry sector influences the $p$ value, we used all selections of the respective industry sectors as sample size for the $\chi^2$ tests. Finally, because we tested a single hypothesis using multiple tests, we used a *Bonferroni correction* to adjust the significance level by dividing the given significance level of 0.05 by the number of tests (the Bonferroni correction is used when several statistical tests are performed simultaneously, which requires an adjustment of the $\alpha$ value[44]). Including the option "Other" in question D005, we provided 20 industry sectors to choose from, that is, the corrected significance level is $p_{B_{cor}} \leq 0.05/20 = 0.0025$.

### 3.3.3 Quantitative analysis for process combinations

To derive process combinations from the data, we analyzed the (combined) occurrence of frameworks, methods, and practices in the dataset. For this analysis, we used the questions PU09 (frameworks and methods), PU10 (practices), and PU04 (combined process use). Our definition of *methods* and *practices* is grounded in the definitions for these concepts as provided by Diebold and Zehler.[45] They define an (agile) method as a process combination addressing the whole software lifecycle, while an (agile) practice is defined as an established instruction with a specific focus. We adopted these definitions such that they are applicable also to traditional and generic methods (frameworks) and practices. To structure the analysis, we defined the *analysis model* shown in Figure 3.

The analysis was performed in multiple steps, and each step was performed twice: (i) on the entire dataset and (ii) on a subset created from a filter using the participants' selection of question PU04, that is, the selection if participants combine different methods intentionally or not. Specifically, the following main analysis steps using the R-package a priori (online: https://www.rdocumentation.org/packages/arules/versions/1.6-2/ topics/apriori) that, among other features, allows for setting recurrence thresholds and minimum/maximum set sizes, were performed:

**Methods:** First, the combined use of the different frameworks and methods, for example, Waterfall, DevOps, and Kanban, was analyzed, and a Top-10-like list of methods and method combinations was computed. The combinations were computed using a recurrence threshold of 35%; that is, we included methods and combinations that were selected by at least 35% of the participants. The recurrence threshold was set to 35% as it identifies a minimal group of three frameworks and methods in the entire dataset and a minimal group of four in the projected dataset generated through PU04 = "Yes."

**Practices:** Similarly, we analyzed the practices, for example, Coding Standards, Code Reviews, and Release Planning. Different to the analysis of the frameworks and methods, we used an 85% recurrence threshold as this threshold provides a minimal group of two practices in the entire dataset as well as in the projected dataset generated through PU04 = "Yes."

**Process variants:** Finally, one exemplary selected hybrid method (i.e., Scrum, Iterative Development, and Lean in the filtered dataset generated through PU04 = "Yes") was further investigated to qualify the process variants. This particular hybrid was selected as a demonstrator for the analysis procedure as having the highest number of process variants. To this end, for each size of combinations, all combinations of practices reaching 85% recurrence threshold within the hybrid were sorted based on the recurrence threshold, and the set of practices sorted based on their first appearance index.

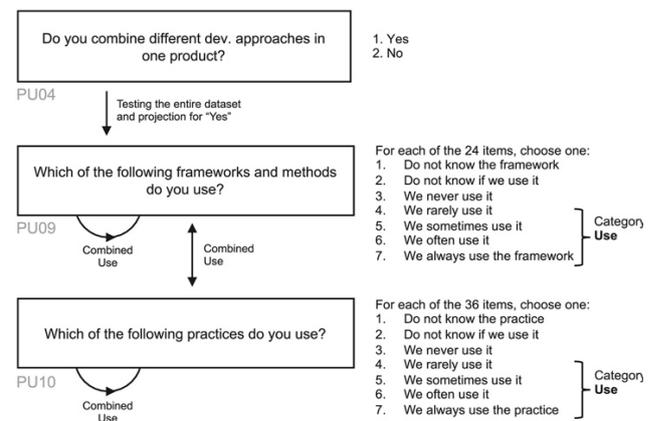

**Figure 3.** Overview of the analysis model used in this study

### 3.4 Validity procedures

To improve the validity and to mitigate risks, we implemented different measures focused around replicability and consistency as well as bias. First, our research is grounded in previously conducted studies. Notably, the key question of this study was derived from the outcomes of our previously conducted study.[10] An extended design team developed the survey instrument as described in Kuhrmann et al.[17] The data analysis was performed by different teams; that is, one team performed the hypothesis testing while another team focused on the quantitative analyses. Researchers not involved in the data analysis were tasked to provide the quality assurance.

**Figure 2.** Overview of the research method applied including this study's position in the overall HELENA project and its previously published outcomes. The figure also includes a world map that highlights all 55 countries that contributed to the HELENA dataset

Second, as one of the main goals of this study is to build a quantitative basis, we opted for the *convenience sampling strategy*[43] to collect the data by accepting the risk of losing full control in terms of sampling, response rate, and so forth. This decision was made to collect as many data points as possible. To handle this risk, before analyzing the data, we implement rigorous data pre-processing including a consistency check of the data (see Section 3.3.1).

## 4. RESULTS

In this section, we present the results of our study. The presentation of the results is structured following the research questions as provided in Section 3.1 and our analysis model as shown in Figure 3.

### 4.1 Checking the base assumptions
As outlined in Section 3.3.2, our study is built on previously published studies[10,15] that found no evidence that the use of combined processes in practice depends on company size or industry sector. Therefore, we tested two hypotheses for which the results are presented in Table 2 (H1; from Kuhrmann et al[10]) and in Table 3 (H2; according to Kuhrmann et al[15]).

Table 3 shows the ratios of participants that do not combine (*non-hybrid*, NH) and those that combine processes (*hybrid*, H) within an industry sector and for all remaining industry sectors. The table shows the individual test results, which, however, have to be considered in the context of the Bonferroni-corrected significance level of $p_{Bcor} \leq 0.0025$ (Section 3.3.3). The results shown in Tables 2 and 3 support the findings from Kuhrmann et al.[10,15] Notably, the results from Table 3—given the Bonferroni correction—show that no $\chi^2$ test is significant, which does not allow for concluding that the industry sector influences the use of hybrid methods. Hence, the results show that the combined use of different frameworks, methods, and practices, that is, the use of hybrid methods, is a common practice in industry. The question for *"What do such combinations look like?"* has therefore to be considered of high relevance.

> **Finding 1:** The use of hybrid development methods has not shown any dependence with regards to either the company size (H1) or the industry sector (H2). Therefore, given the high p-value of the majority of the tests, the use of hybrid development methods can be considered state of practice across companies of all sizes and in all industry sectors.

### 4.2 Combined use of frameworks and methods
The first step in the quantitative analysis is the investigation of the combined use of frameworks and methods. Of the 1467 data points, 845 provide data on the use of frameworks and methods, that is, answers to question PU09 (Figure 3). As shown in Figure 3, this multiple-choice question provided 24 items to choose from complemented with a free-text option. Of the 845 data points, 792 had multiple selections. The quantitive analysis of the combined use of frameworks and methods is performed twice: once for all data points and again for those data points for which the study participants stated to use hybrid methods intentionally.

#### 4.2.1 Analysis of all data points
Figure 4 (left) shows the resulting combinations using the 35% threshold for the combined process use in the entire (non-filtered) dataset. This threshold results in 17 groups of two or three combined frameworks and methods—there is no group with four or more elements with at least 35% agreement regarding the combined process use. *Scrum* is the most frequently selected method (674 participants), which is followed by *Iterative Development* (620) and *Kanban* (523). Extending the scope to framework and method combinations, a number of couples and *all* triplets include *Scrum*. Expected combinations are present, for example, *(Scrum–Kanban–DevOps)*, which was stated by 309 participants, and the "Water-Scrum-Fall," that is, *(Waterfall–Scrum)*, according to West et al,9 which was mentioned by 380 participants. Please note that, in the following, we consistently present clusters of frameworks, methods, and practices in the form $(Item_1 - ... - Item_n)$ to make clear, which $n$ frameworks, methods, and practices together form a cluster, that is, a group that can be combined with further individual frameworks, methods, practices, and even other clusters.

#### 4.2.2 Analysis of intentionally combining practitioners
Applying the explicitly stated combinations of methods and practices, that is, question PU04 (Figure 3) as a filter, that is, re-running the analysis for only those participants that explicitly claimed to use the different frameworks, methods, and practices in combination, Figure 4 (right) results in 27 groups of two to four explicitly combined frameworks and methods, whereas there is no group with five elements or more having at least 35% agreement regarding combined process use. The combined frameworks and methods as shown in Figure 4 *do not* provide the full picture as they only form the "core," but are complemented with further frameworks, methods, and practices, which will be elaborated in more detail in the following sections.

> **Finding 2:** Among the 24 frameworks and methods presented to the study participants, we identified 17 (entire dataset, Figure 4, left) and 27 (dataset filtered for question PU04, Figure 4, right) core groups with two to four elements for which the study participants agree with at least 35% on their combined use. These combinations are based on eight base methods that provide the frame for hybrid development methods.

### 4.3 Combined use of frameworks, methods, and practices
The second step in the quantitative analysis is the investigation of the combined use of frameworks, methods, *and* practices (Figure 3, PU09 and PU10). Of the 1467 data points, 769 provide data. As shown in Figure 3, questions PU09 and PU10 provided 36 items each to choose from and a free-text option. For each of the 36 items, we used a Likert-scale to rate the use *and* the frequency of use. The following analyses are based on those answers that we categorized into the category "Use" (Figure 3). In total, we used 769 data points for analysis of which 742 had multiple selections.

**Table 2.** Result of testing H10: The use of hybrid methods does not depend on the company size

| Id | Results | Decision | Kuhrmann et al.[10] |
|---|---|---|---|
| H1$_0$ | $\chi^2$ = 1.9972, df = 3, p = 0.573 | No support | No support |

**Table 3.** Results for testing $H2_0$: The use of hybrid methods does not depend on the industry target domain (with corrected significance level $p_{B_{cor}} \leq 0.0025$ for 20 industry target domain options; NH: not hybrid, i.e., PU04 = "No"; H: hybrid, i.e., PU04 = "Yes")

| Industry sector | This sector NH:H | Other sectors NH:H | $\chi^2$ | p value |
|---|---|---|---|---|
| Automotive Software and Systems | 14 : 66 | 163 : 515 | 1.37 | 0.24 |
| Aviation | 09 : 21 | 168 : 560 | 0.43 | 0.51 |
| Cloud Application and Services | 29 : 95 | 148 : 486 | 0.00 | 1.00 |
| Defense Systems | 02 : 26 | 175 : 555 | 3.38 | 0.07 |
| Energy | 07 : 30 | 170 : 551 | 0.21 | 0.65 |
| Financial Services | 34 : 149 | 143 : 432 | 2.73 | 0.10 |
| Games | 01 : 17 | 176 : 564 | 2.32 | 0.13 |
| Home Automation and Smart Buildings | 05 : 17 | 172 : 564 | 0.00 | 1.00 |
| Logistics and Transportation | 11 : 43 | 166 : 538 | 0.14 | 0.71 |
| Media and Entertainment | 06 : 25 | 171 : 556 | 0.10 | 0.75 |
| Medical Devices and Health Care | 13 : 61 | 164 : 520 | 1.20 | 0.27 |
| Mobile Applications | 19 : 105 | 158 : 476 | 4.82 | 0.03 |
| Other Embedded Systems and Services | 09 : 46 | 168 : 535 | 1.22 | 0.27 |
| Other Information Systems | 20 : 87 | 157 : 494 | 1.22 | 0.27 |
| Public Sector and Contracting | 21 : 72 | 156 : 509 | 0.00 | 0.95 |
| Robotics | 01 : 17 | 176 : 564 | 2.32 | 0.13 |
| Space Systems | 08 : 26 | 169 : 555 | 0.00 | 1.00 |
| Telecommunication | 07 : 38 | 170 : 543 | 1.19 | 0.27 |
| Web Applications and Services | 40 : 162 | 137 : 419 | 1.68 | 0.20 |
| Other | 27 : 63 | 150 : 518 | 2.12 | 0.15 |

As described in Section 3.3.3, to analyze the combinations of practices within the base methods and method combinations provided in Section 4.2, we used an 85% threshold for the agreement regarding the combined use. That is, for each method combination identified in Section 4.2, the combinations of practices within these have been computed. All analyses were performed using base methods and method combinations resulting from the entire dataset and from the filtered dataset based on the answers to the question PU04 (Figure 3). The overall result is shown in Figure 6, which will be described step by step in the following subsections.

### 4.3.1 Unfiltered practices

As a first step, the (non-filtered) dataset was analyzed for the most commonly used practices, that is, those practices with the highest agreement regarding combined use without a particular combination of methods. To find these practice combinations and to find those groups that have the largest agreement in the entire dataset, we explored the dataset. The smallest group with the highest agreement in the data was the pair *Code Review* and *Coding Standards* ($n = 674$, agreement = 0.87). The agreement level of 0.87 was also used to set the threshold of 85% agreement as introduced in Section 3.3.3.

The results of the analyses of the entire (non-filtered) dataset using the 85% threshold are shown in Figure 5. The figure shows for the entire dataset three practices (*Code Review*, *Coding Standards*, and *Release Planning* with an 85% agreement) for which one pair of two practices with an 85% agreement could be find. Likewise, in the filtered dataset (after applying PU04 as filter), five practices could be identified (*Code Review*, *Coding Standards*, *Release Planning*, *Automated Unit Testing*, and *Protoyping* with an 85% agreement). Of these five practices, three pairs of two com- posed from the five practices could be identified, which have at least 85% agreement among the participants of the study.

### 4.3.2 Individual practices

In the same reading as for Figure 5, the upper part of Figure 6 presents the practices reaching 85% agreement within the context of the respective base methods and method combinations. The upper-left part of Figure 6 presents the results for the entire (non-filtered) dataset, while the upper-right part of Figure 6 presents the results for those base methods and method combinations computed from the filtered dataset after applying PU04 as a filter (Section 4.2).

For each practice (Figure 3; PU10, 36 items to choose from), Figure 6 shows the assignment to a base method or a method combination for which 85% agreement could be found in the dataset. The total number of such practices assigned to a particular method combination is shown in the row "Number of practices in combinations" beneath the respective method combinations. For instance, for the method combination *(Scrum- Kanban)*, 14 practices are assigned to this method combination in the entire dataset, and, respectively, 15 practices are assigned to this method

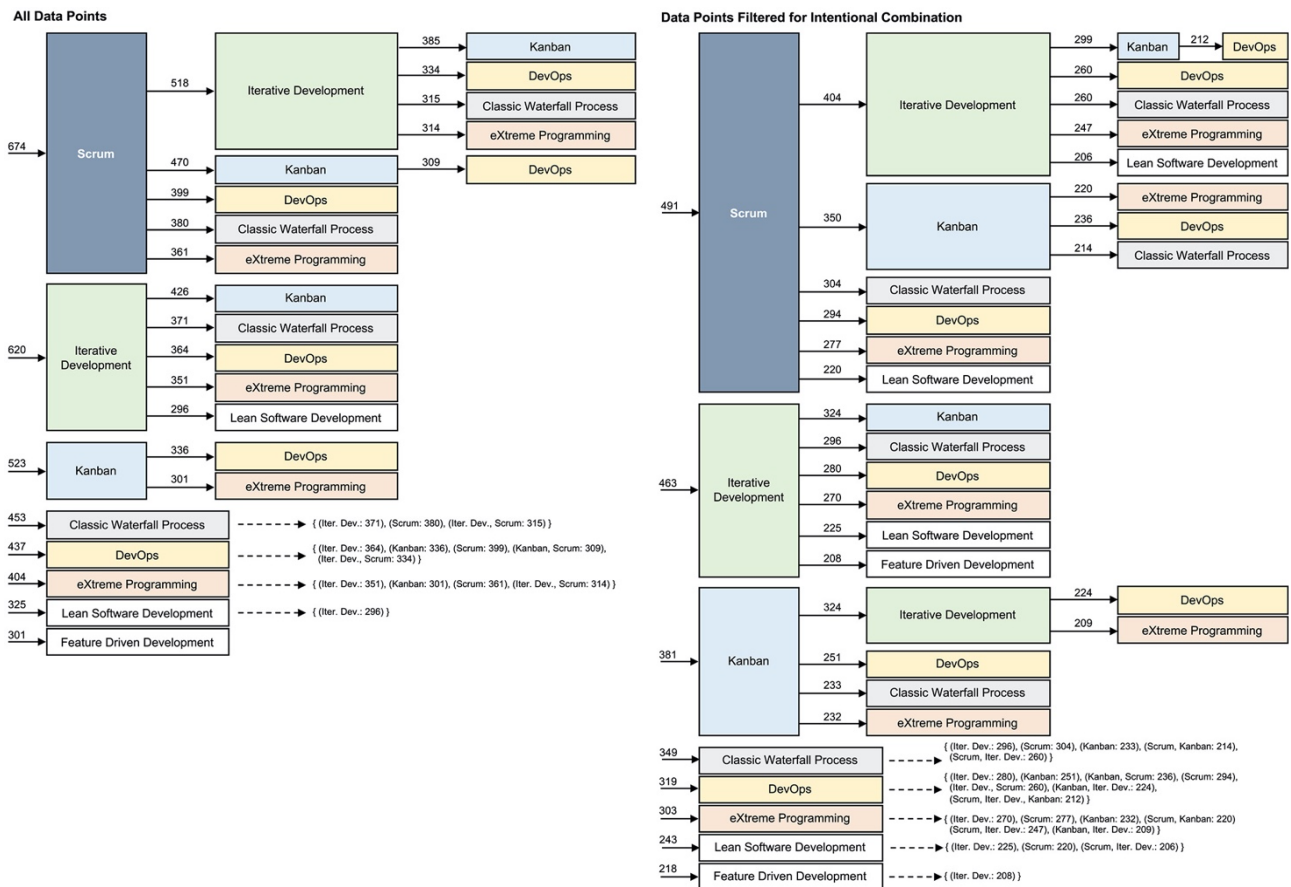

**Figure 4.** Base methods and method combinations (35% threshold) in the non-filtered dataset (left) and on the dataset filtered for intentional combination of different development frameworks and methods (right). The figures reads from the right to the left, e.g., 309 participants use Scrum, Kanban and DevOps in combination (all data points). The lower part of the figure (dashed arrows) shows the other/alternative entry points for those combinations that are part of the combinations shown in the upper part of the figure

combination in the PU04-filtered dataset. All possible combinations of frameworks, methods, and practices with 85% agreement are constructed from these individual practices, which is elaborated in more detail in Section 4.3.3. The visualization in the upper part of Figure 6 allows for two main observations:

**First Observation:** The sparsity of rows and thus a limited number of practices consistently selected by the participants in the context of a given method or method combination.

**Second Observation:** The selected practices (highlighted rows) are consistent across different method combinations. That is, a limited number of practices is consistently used with an agreement of at least 85% regardless of the actual method combination.

In addition, two minor observations can be made: first, the "density" of the practices for which the participants agree regarding their combined use is higher in the PU04-filtered dataset than in the non-filtered dataset, that is, in that share of the data in which the participants explicitly stated to combine multiple frameworks, methods, and practices. Second, it seems that as if the larger the number of combined methods is the more practices find an agreement among the participants. For instance, the rightmost method combination in Figure 6 *(Scrum-Iterative Development-Kanban-DevOps)* has 21 practices assigned for which the participants find an agreement of at least 85%.

### 4.3.3 Combinations of practices

The lower part of Figure 6 extends the analysis from Section 4.3.2 starting with the "Number of practices in combinations" row. This row shows how many practices are assigned to the different method combinations, thus forming the basis for framework, method, and practice combinations to derive hybrid development methods. Within these sets of practices, we search for *practice tuples* of increasing size having an agreement of at least 85% and that are used in combination in the respective method combination.

Taking the combination *(Scrum-Kanban)* as an example, 14 (entire dataset) and 15 (PU04-filtered dataset) practices are assigned to this combi- nation. In the first step, we search

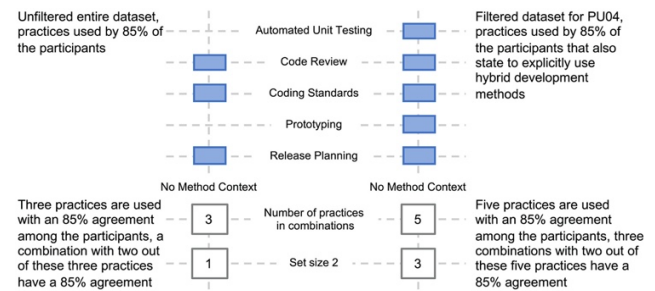

**Figure 5.** Overview of the most frequently used practices in the entire dataset (left) and in the filtered dataset (PU04, right). The figure illustrates the most frequently used practices and also shows how many possible combinations can be found with 85% agreement among the participants

for pairs of two practices from these 14 (15) practices with the required agreement level. This results in 48 pairs (for the entire dataset) and, respectively, 65 pairs (PU04-filtered dataset) of practices out of 14 (15) practices. In the next step, we search for 3-tuples, then for 4-tuples, and so forth until no *x*-tuple with the required agreement level is found. As Figure 6 shows, the biggest set size with an agreement of at least 85% is eight. For instance, in the PU04-filtered dataset and for the combination *(Scrum-Iterative Development-Kanban- DevOps)*, three 8-tuples of practices from the 21 practices assigned to this combination can be found in the dataset. Hence, the lower part of Figure 6 allows for two main observations:

**First Observation:** The larger the combination size, i.e., the more practices are included in the set of practices individually reaching the required 85% agreement, the more "general" agreement can be found regarding the practices that would be included in a method or method combination.

**Second Observation:** The larger the number of combinations within a group the more practices are consistently selected by the participants. This provides a quantitative description of the process variants within a specific hybrid method.

Also, similar to the observations from Section 4.3.2, we see that the larger the number of combined methods, the more combinations of practices find agreement among the practitioners, and the size and the amount of combinations are bigger in the PU04-filtered data than in the entire dataset.

**Finding 3:** In summary, analyzing the 36 practices and their relation to the methods and method combinations found in Section 4.2, we find few practices only that find agreement of at least 85% among practitioners. However, as shown in Figure 6 the assignments show a consistency across the base methods and method combinations. That is, few practices only are consistently used for hybrid methods, yet, allow for framing a large number of process variants.

### 4.4 Investigating process variants

The clusters of frameworks, methods, and practices as presented in Figure 5 and Figure 6 (Section 4.3.3) only show *which* frameworks, methods, and practices are potentially combined with one another. However, all practices involved in theses analyses reached an agreement level of at least 85%, and thus, it is necessary to provide a better characterization of when a particular practice is included in a combination, that is, to rank the individual practices.

For this, we "re-apply" our data analysis procedure to determine the practices' ranks. Figure 7 outlines our approach, which is discussed in detail in this section. To showcase our approach, we selected the method combination that yielded the highest number of potentially relevant combinations, that is, the method combination *(Scrum-Iterative Development-Lean Software Development)* for which up to 643 process variants can be constructed. Having selected this method combination, we apply the following procedure (as illustrated in Figure 7):

1. Filter the dataset to isolate those data points related to the hybrid method under investigation (i.e., *(Scrum-Iterative Development-Lean)* in our case) and for which participants answered PU04 with "Yes."

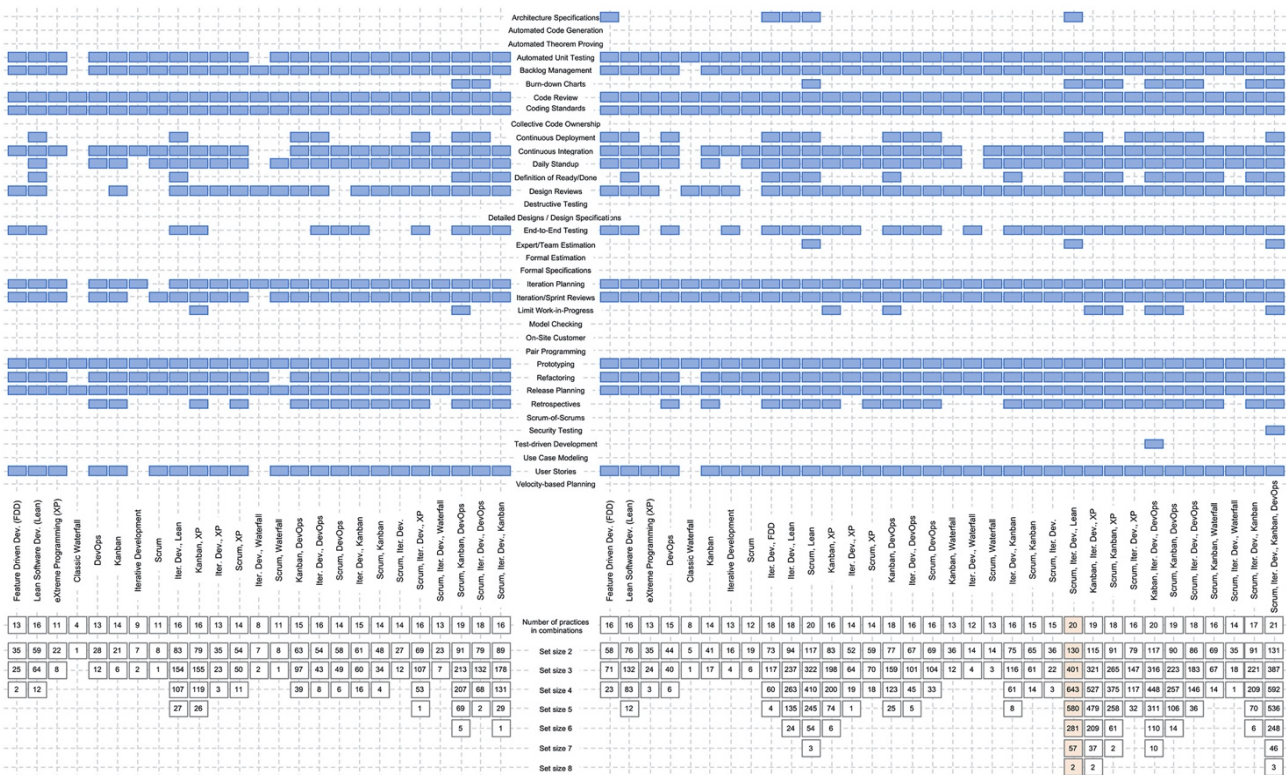

**Figure 6.** Overview of the practices used with 85% agreement in hybrid development methods. The left part of the figure shows the analysis results for the entire, non-filtered dataset, the right part illustrates the analysis results for the filtered dataset based on the value of question PU04. The upper part of the figure shows the practices used together with the different method combinations, and the lower part shows the possible combinations of practices of given set sizes within the different method combinations, based on the *number of practices in combinations*-row

2. Identify the combinations, specifically, the combinations with their growing set sizes (Figure 6 lower part).

3. For each present combination and set size (i.e., 1 to 8 in our case) do:
   (a) Identify the combinations with a recurrence threshold of at least 85%.
   (b) Sort the set based on recurrence threshold.
   (c) Starting from the combination with the highest recurrence threshold, iterate on the sorted set to identify any new practice and associate to the practice the index of the combination.

Figure 8 visualizes the results of the analysis. In the figure, rows are internally ordered based on the sorting established for set size 4, which is the combinations size with the highest number of variants (i.e., 643). For each set size, the first row reports the index of the first variant in which a given practice is detected, while the second row indicates the agreement on the process variant within the data set. That is, for the selected case, the practices *Code Review*, *Coding Standards*, *Refactoring*, and *Release Planning* appear "immediately" just in the starting configuration. Just in the next process variant, *Backlog Management* is added; in the next one, *Prototyping* is added, and so forth. After *Iteration Planning*, in this combi- nation, there are six more combinations of the up to seven practices found so far, before *Iteration/Sprint Planning* is added as new practice in the 11th combination. Likewise, 260 combinations including up to 15 practices appear in the ranked list of practices before the *Definition of Done/- Ready* appears for the first time. From this, we can conclude the relevance of a specific practice in a hybrid method. For instance, in sets of size 4, we can see that *Burn-down Charts*, *Continuous Deployment*, and *Expert/Team-based Estimation* are practices mentioned only in combinations at the end of the sorted set (i.e., indexes of 489, 490, and 563 on a set with 643 process variants), which makes them less relevant for constructing a hybrid method in this context.

Looking at the highlighted row of Figure 8, it can be observed how some practices are much more appealing to the participants as occurring within the variants that have the highest recurrence threshold, which leads to the following two observations:

**First Observation:** Practices on the left side are elements that characterize the hybrid process variant under investigation, while practices on the right side are less relevant in the hybrid method.

**Second Observation:** Even though all rows in Figure 8 follow the ordering derived from the variants in the set of size 4, the sorting on the other set sizes is rather accurate and aligned.

Hence, Figure 8 does not only show which frameworks, methods, and practices are combined, but also the "preferred" practices used to embody the base methods and/or the method combinations framing hybrid methods as identified in Section 4.2. Even with the high level of agreement of 85% regarding the combined use, Figure 8 shows that there are practices with a higher agreement among the practitioners, and these agreement levels together with the rank of a practice (the index of its first appearance in the sorted list) indicate the more "prominent" hybrid methods. On the other hand, the figure also shows that the increasing size of the practice cluster does not necessarily lead to a high variability (see set size 8, which only has two variants that are characterized by the presence of *Backlog Management* or the lack thereof).

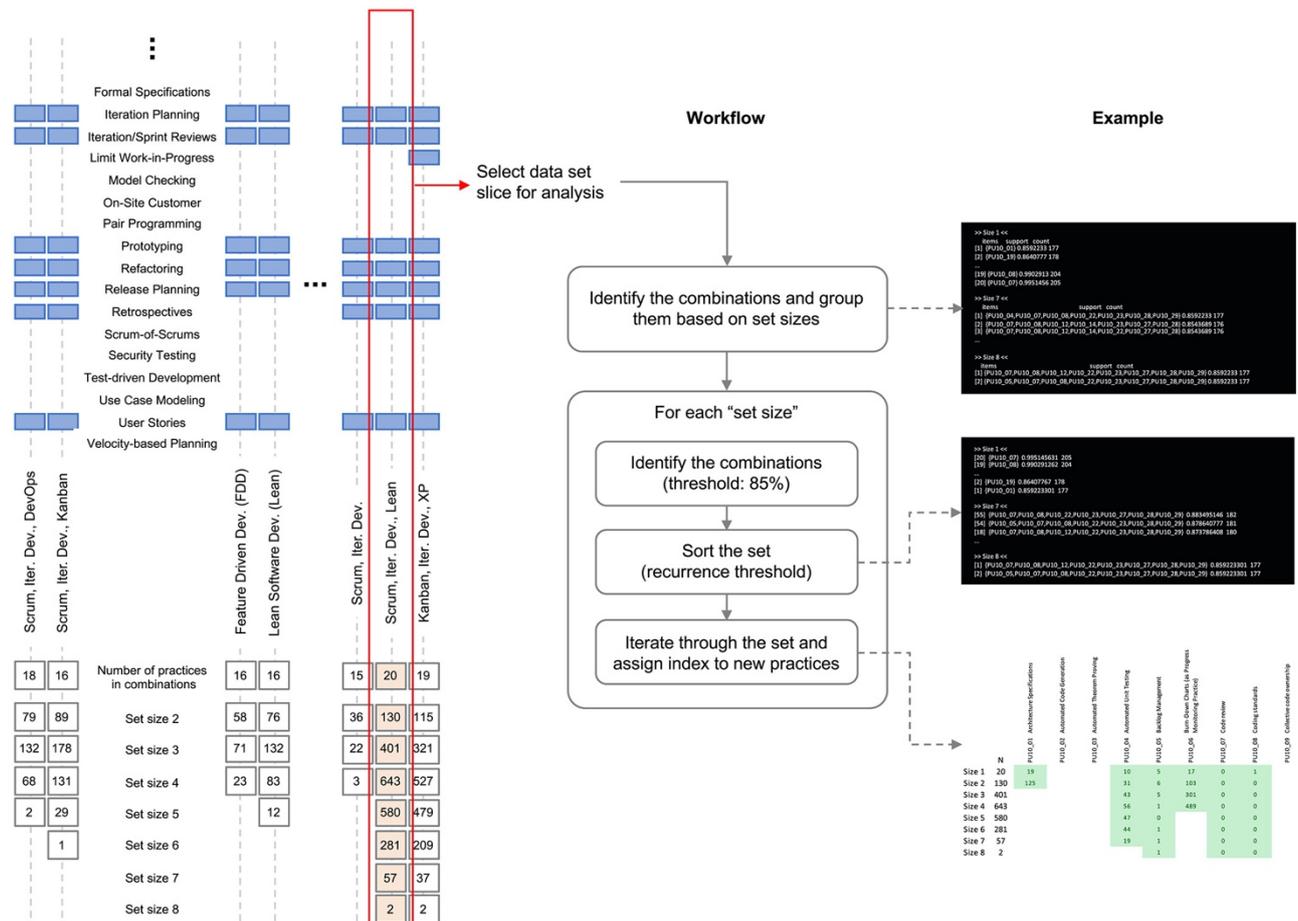

**Figure 7.** Illustration of the basic procedure to analyze and rank the different process variants. The figure also shows the selected hybrid method from Figure 6 for which we showcase the analysis procedure

| | Code review | Coding standards | Refactoring | Release planning | Backlog Management | Prototyping | Iteration Planning | Iteration/Sprint Reviews | Daily Standup | Design Reviews | Continuous integration | Automated Unit Testing | End-to-End (System) Testing | User Stories (as Requirements Engineering Practice) | Retrospectives | Definition of done / ready | Burn-Down Charts (as Progress Monitoring Practice) | Continuous deployment | Expert/Team based estimation (e.g. Planning Poker) | Architecture Specifications |
|---|---|---|---|---|---|---|---|---|---|---|---|---|---|---|---|---|---|---|---|---|
| Set size 1 20 Process Variants | 0 0,995 | 1 0,990 | 3 0,976 | 2 0,976 | 5 0,966 | 4 0,971 | 6 0,961 | 7 0,956 | 11 0,937 | 9 0,942 | 8 0,947 | 10 0,937 | 12 0,927 | 13 0,922 | 14 0,898 | 15 0,898 | 17 0,874 | 16 0,888 | 18 0,864 | 19 0,859 |
| Set size 2 130 Process Variants | 0 0,985 | 0 0,985 | 2 0,971 | 1 0,971 | 6 0,961 | 4 0,966 | 9 0,956 | 11 0,951 | 24 0,937 | 25 0,937 | 19 0,942 | 31 0,932 | 34 0,927 | 44 0,917 | 74 0,893 | 75 0,893 | 103 0,869 | 86 0,883 | 112 0,864 | 125 0,854 |
| Set size 3 401 Process Variants | 0 0,966 | 0 0,966 | 0 0,966 | 1 0,961 | 5 0,951 | 2 0,956 | 6 0,947 | 13 0,942 | 25 0,932 | 26 0,932 | 17 0,937 | 43 0,922 | 42 0,922 | 86 0,908 | 197 0,883 | 179 0,888 | 301 0,864 | 219 0,879 | 333 0,859 | |
| **Set size 4 643 Process Variants** | 0 0,951 | 0 0,951 | 0 0,951 | 0 0,951 | 1 0,942 | 2 0,942 | 5 0,932 | 11 0,927 | 16 0,922 | 24 0,917 | 26 0,917 | 56 0,908 | 75 0,903 | 129 0,893 | 260 0,879 | 261 0,879 | 489 0,859 | 490 0,859 | 563 0,854 | |
| Set size 5 580 Process Variants | 0 0,927 | 0 0,927 | 0 0,927 | 0 0,927 | 0 0,927 | 1 0,927 | 2 0,922 | 4 0,913 | 8 0,908 | 10 0,908 | 22 0,903 | 47 0,893 | 67 0,888 | 94 0,883 | 236 0,869 | 237 0,869 | | | | |
| Set size 6 281 Process Variants | 0 0,903 | 0 0,903 | 0 0,903 | 0 0,903 | 1 0,903 | 1 0,903 | 0 0,903 | 0 0,903 | 4 0,893 | 5 0,893 | 23 0,883 | 44 0,874 | 69 0,869 | 99 0,864 | 205 0,854 | 206 0,854 | | | | |
| Set size 7 57 Process Variants | 0 0,883 | 0 0,883 | 0 0,883 | 0 0,883 | 1 0,879 | 0 0,883 | 1 0,883 | 0 0,883 | 2 0,874 | 4 0,874 | 9 0,869 | 19 0,859 | | | | | | | | |
| Set size 8 2 Process Variants | 0 0,859 | 0 0,859 | 0 0,859 | 0 0,859 | 1 0,859 | 0 0,859 | 0 0,859 | 0 0,859 | 0 0,859 | | | | | | | | | | | |

**Figure 8.** Characterization of the variants of the *(Scrum-Iterative Development-Lean Software Development)* hybrid method. Rows are internally ordered based on the sorting established for set size 4, that is, the combinations size with the highest number of variants and the running example used in the text

> **Finding 4:** Analyzing the variance of a specific hybrid method through the procedure outlined in Figure 7, it was possible to identify and sort a small set of practices that characterize the given hybrid method based on the participants' selections. Moreover, no matter the threshold chosen with regards to the agreement level among participants' responses, a noticeable consistency can be seen in the ordering of the practices, which indicates the presence of a core set of practices.

**4.5 Constructing hybrid development methods**
The analyses conducted in the Sections 4.2 to 4.4 provide important insights regarding the *base methods*, the basic *method combinations*, and the *number of practices assigned* to these base methods and method combinations.

In this section, we demonstrate how to utilize our analysis method to incrementally construct hybrid development methods. For this, we apply the following procedure:

1. Based on the smallest groups of frameworks and methods for the entire dataset and for the PU04-filtered dataset shown in Figure 4, we form the "umbrella" to construct our hybrid method. Our current data allows for constructing 17 groups (all data) and, respectively, 27 groups (PU04-filtered data), which are constructed from eight frameworks and methods in total.

2. Based on the smallest groups of practices for the entire dataset and for the PU04-filtered dataset, which are shown in Figure 5, we form the "core" of practices from the different pairs. Our current data allow for constructing one pair for the entire dataset and three for the PU04-filtered dataset.

3. For each base method or method combination created in the first step, we add the core(s) as created in the section step to set an extended method context. For instance, instead of looking for all practices to be combined with *Scrum*, we search all additional practices meeting the required agreement level of 85% for the new combination *(Scrum-corei)* with *i* denoting the cores identified.

4. In the final step, we integrate all frameworks, methods, and practices into hybrid methods by building the unique combinations (the process variants) of all these components. Applying our analysis from Section 4.4, we can read the "promising" variants for our hybrid method. Hence, we can create all process variants and rank them using the participants' level of agreement.

That is, we aim to identify those practices (if any) that are included in bigger combinations containing the "core" for each of the base methods or method combinations. We applied this procedure to both the entire dataset and the PU04-filtered dataset, which results in the *(method-corei-practice)* combinations shown in Figure 9 (left) for the entire dataset and Figure 9 (right) for the PU04-filtered dataset.

**4.5.1 Constructing the "Water-Scrum-Fall"**
Taking the *Classic Waterfall* as an example, we see in both parts of Figure 9 that the *Waterfall* is characterized by the core-practices only. Moving on to "Water-Scrum-Fall," that is, the combination *(Scrum-Waterfall)*, in the entire dataset (Figure 9, left), we see one single combination of size three containing the *(Scrum-Waterfall)* method combination, the core consisting of *(Code Review-Coding Standards)*, and *Release Planning* as third practice. In the PU04-filtered data (Figure 9, right), "Water-Scrum-Fall" is characterized by the *(Scrum-Waterfall)* method combination and three cores of which either is extended by *(Prototyping-Iteration/Sprint Reviews)* as illustrated below:

$$(Scrum\text{-}Waterfall) + \begin{bmatrix} (Code\ Review\text{-}Coding\ Standards) \\ (Code\ Review\text{-}Release\ Planning) \\ (Coding\ Standards\text{-}Release\ Planning) \end{bmatrix}_{XOR} + (Prototyping\text{-}Iteration/Sprint\ Reviews)$$

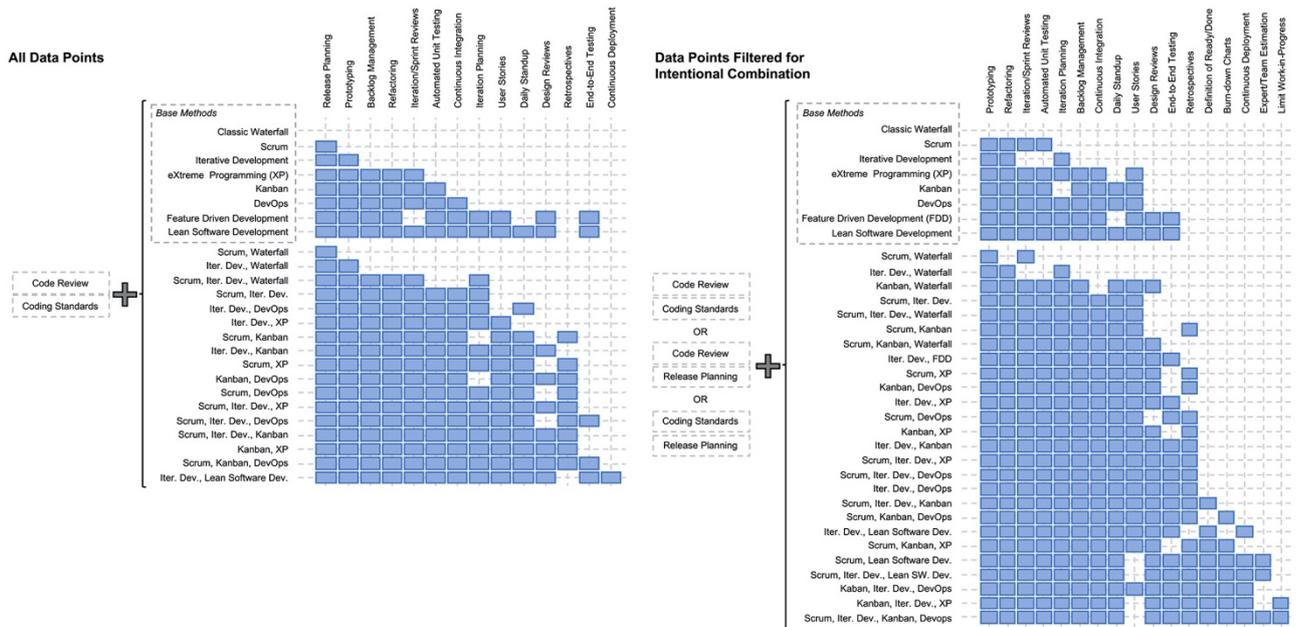

**Figure 9.** Combinations resulting from applying the statistical analysis method: The right part of the figure characterizes the base methods (top) and method combinations (bottom) generated by identifying recurring combinations of three practices with at least 85% agreement level that contain the core of the entire non-filtered dataset, that is, (*Code Review-Coding Standards*). The right part of the figure shows a refined characterization using the dataset filtered for question PU04 of base methods (top) and method combinations (bottom). The three cores used are: (*Code Review-Coding Standards*), (*Coding Standards-Release Planning*), (*Code Review-Release Planning*)

Both parts of Figure 9 also show the gradual increase of the practice pools. Staying with *Scrum* in the entire dataset, that is, (*Scrum- Code Review-Coding Standards-Release Planning*), the base method *Iterative Development* "extends" this combination with *Prototyping*, that is, the combination would be (*Iterative Development-Code Review-Coding Standards-Release Planning-Protyping*). That is, *Scrum* is characterized by the practice combination (*Code Review-Coding Standards-Release Planning*) and *Iterative Development* is characterized by (*Code Review-Coding Standards-Release Planning-Prototyping*).

### 4.5.2 Constructing a hybrid method from the variant space

As Figure 9 only shows which frameworks, methods, and practices to combine, we use our findings from Section 4.4 to provide a recommendation, based on the practices' rank. Below, the first quadruple from Figure 8 is outlined:

$$(Scrum-Iterative\ Development-Lean\ SWDev.) + \begin{bmatrix} Code\ Review \\ Coding\ Standards \\ Refactoring \\ Release\ Planning \end{bmatrix}_{a_4=0.951}^{All}$$

It can be seen that the overall agreement level for this practice cluster is $a_4 = 0.951$, where $a_4$ points to the "initial" cluster of practices with a set size of 4 from Figure 8. Moving on to the first index in Figure 8, *Backlog Management* is added. To keep the quadruples, that is, implementing the requirement of including four practices only in our hybrid method, instead of just reusing the set above, in this case, we have to limit the selection accordingly to three, allowing *Backlog Management* to be added as forth:

$$(Scrum-Iterative\ Development-Lean\ SWDev.) + \begin{bmatrix} Code\ Review_{a_3=0.966} \\ Coding\ Standards_{a_3=0.966} \\ Refactoring_{a_3=0.966} \\ Release\ Planning_{a_3=0.961} \end{bmatrix}_{a_4\to 3=[0.951,0.966]}^{3\ out\ of\ 4} + (Backlog\ Management_{a_4=0.942})$$

This has several consequences: first, instead of quadruples, we now have to build *triplets* from the four available practices. That is, for the practices, we have to move from the set size 4 to the set size 3 (Figure 8), which impacts the level of agreement. To make this visible, the agreement level is added to the practices, and $a_3$ points to the set size 3. Second, the boundaries for the agreement level change, which is shown by the index $a_4 \to a_3 = [0.951, 0.966]$, where the lower bound is taken from the lowest $a_4$-agreement level and the upper bound is taken from the highest $a_3$-agreement level.

Yet, this is the first selection step only. The next consequence is the necessity for re-computing the specific agreement levels for the resulting triplets. Therefore, the algorithm described in Section 4.4 needs to be re-executed for the triplets. In total, we can construct four triplets from the four given practices. The result of the algorithm's execution is shown in Table 4. The outcome, that is, the minimal level of agreement is now added to the formal representation ($a_3 \geq 0.956$), which means that every triplet has at least an agreement level of 95.6%.

The next step is now adding the practice *Backlog Management*. We are going back to four practices; that is, we have to re-compute the agreement levels with quadruples again, but, instead of using the "initial" four practices, we now used the triplets and the practice *Backlog Management*. Re-computing the new quadruples results in a minimal level of agreement $a_4 = 0.932038835$ as shown in Table 5. That is, selecting one triplet from the "initial" four practices and adding *Backlog Management* as a fourth

Table 4. Results for re-computing the agreement levels for triplets in the construction of a hybrid method

| Combination | Code review PU10_07 | Coding standards PU10_08 | Refactoring PU10_28 | Release planning PU10_29 | Agreement for $a_3$ |
|---|---|---|---|---|---|
| Combination 1 | X | X | X | | 0.966019417 |
| Combination 2 | X | X | | X | 0.961165049 |
| Combination 3 | | X | X | X | 0.95631068 |
| Combination 4 | X | | X | X | 0.95631068 |

*Note*: The computed minimal level of agreement for three out of four practices is $a_3$ = 0.95631068.

Table 5. Results for re-computing the agreement levels for quadruples in the construction of a hybrid method

| Combination | Code review PU10_07 | Coding standards PU10_08 | Refactoring PU10_28 | Release planning PU10_29 | Agreement for $a_3$ |
|---|---|---|---|---|---|
| Combination 1 | X | X | X | | 0.966019417 |
| Combination 2 | X | X | | X | 0.961165049 |
| Combination 3 | | X | X | X | 0.95631068 |
| Combination 4 | X | | X | X | 0.95631068 |

*Note*: The computed minimal level of agreement for three out of four practices and *Backlog Management* is $a_4$ = 0.932038835.

practice results into an overall minimal agreement level of 93.2%.

Having executed the different computations, the final hybrid method can be constructed as shown below: The starting point is the method combination (*Scrum-Iterative Development-Lean Software Development*), which is complemented by three out of the four initial practices that provide a minimal agreement level of $a_3 \geq 0.956$ and, finally, the practice *Backlog Management* is added, which results into a final minimal agreement level of $a_4 \geq 0.932$.

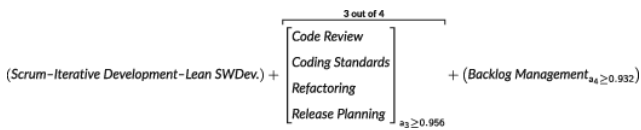

**Finding 5:** Applying our analysis procedures, we can define a statistical construction procedure for describing hybrid methods. Also, our description is not only limited to specific hybrid method instances; we can also characterize a hybrid method and its process variants.

## 5. DISCUSSION

Having presented our results, we conclude this paper by answering the research questions, discussing our findings and discussing the threats to validity.

### 5.1 Answering the research questions

The findings provide a rich quantitative basis and evidence to answer our research questions posed in Section 3.1:

**RQ1:** Klünder et al.[16] found that processes mainly evolve into hybrid methods and provided evidence and generalized the claim by West et al.[9] about the "Water-Scrum-Fall." With this paper, we provide insights regarding eight base methods that are recurrently combined to form hybrid development methods (cf. Finding 2).

**RQ2:** In this paper, we identify the most frequently used practices and how these are combined with each other. Our results reveal a small core of practices used by practitioners regardless of the (hybrid) development method selected (cf. Finding 3).

**RQ3:** Analyzing the variability within the hybrid development methods identified, it is possible to identify small sets of practices that characterize each hybrid method. These practices are consistently mentioned in relation to the hybrid, indicating process variants that are clearly preferred over others (cf. Finding 4).

**RQ4:** In modern software and system development, methods and practices are often combined into processes that are context-dependent. How- ever, when attempting to characterize the different methods by systematically constructing the set of practices using a bottom-up strategy, we show that the resulting combinations of practices vary very little and consistently repeat the same practices (cf. Finding 5).

### 5.2 Hyped methods and old practices

So far, we could identify eight base methods and few practices that, together, are the heart of hybrid development methods. Figure 9 shows these key components that find agreement across the participants of the HELENA study. Looking closer at the practices, we see that hybrid methods are heavily composed of (mostly technical) practices that have been used in software development for decades, namely, *Code Review*, *Coding Standards*, and *Release Planning*. On the one hand, the sets of practices and their assignment to methods show obvious similarities as shown in Figure 6. On the other hand, it is not trivial to identify a specific method or method combination from the practices used in a specific context.

Accepting that formally defined methods are not applied in practice[46-48] and that hybrid development methods are the norm[9-11] (cf. Section 4.1), we can pose a number of further questions. From our perspective, the most urgent question is for the actual role of methods, for instance: What is the value of stating that someone, for example, uses Scrum, when no one does it by the book? What are the implications for devising a particular method if, in practice, the method starts evolving[16] into a hybrid? What are the implications for software process improvement (SPI) programs if, regardless of the "showroom method" providing the umbrella, the practices are the *stable* key components of organizing and conducting the actual project work? What are the implications for educators if companies require students trained in latest methods,[49] but key is a solid understanding of the "old stuff," which is still the core of modern software engineering practice?

Brooks[50] argued that there is no *"Silver Bullet"* that would fit all the different flavors that the software engineering industry has. However, especially in the last decade, it appears that Brooks' observation has been forgotten, and some methods are relentlessly advertised as *the* silver bullet. *New* methods are continuously spawned, and people engage in rather unhealthy discussions arguing whether one method is superior over another. So, are we chasing the white rabbit by focussing on hyped methods and substituting agility with Scrum? One could disagree as "revolutions" like the Agile Manifesto changed the industry inside out, which is certainly true and industry did progress. Yet, such revolutions changed the mindset and the culture of companies, *not* the practices. Test-driven development, continuous integration, continuous delivery, continuous deployment are incarnations of building blocks that already existed, but, in Beck's words[2], have been "cranked up all the knobs to 10." What has changed is how they are combined and how they are used. We argue that we should distance ourselves from such discussions about the "right" method, but should focus our attention to the practices. Studying the nuances behind practices and their implementation in different contexts would possibly lead to interesting findings whether some hybrid methods (sets of practices) are more effective than others.

As shown in this paper, among practitioners, strong agreement can be found at the practice level, and, when analyzing at this level of granularity, methods and frameworks fade into the background. Therefore, we argue that *researchers* should report on the actual practices when presenting cases, as assumptions on practices used based on a method or framework do not hold, *practitioners* should be mindful about new hypes as the identified core practices are building blocks that are agnostic of methods and, finally, *educators* should put more emphasis on teaching practices rather then methods by explaining the rationale behind them and the different ways in which they can be executed.

**5.3 Threats to validity**

In this section, we discuss the threats to validity according to the schema provided by Wohlin et al[51] and in relation to the constructive validity procedures as outlined in Section 3.4.

**5.3.1 Construct validity**

First, as we used an online questionnaire, the options provided in the questions might have both been incomplete and prone to misunderstandings resulting in potentially incomplete or wrong answers. Several tactics have been taken to reduce and mitigate these threats. Notably, multiple-choice questions were complemented with a free-text option for potentially necessary clarification; the questionnaire was made available in four different language to lower the risk of misunderstanding due to language issue (translation performed by native speakers); the questionnaire was constructed by a team of researchers, which tested and revised it (see Section 3.2); and the questionnaire was first released in Europe as a further quality controll,10 which led to final revision. Second, the tactics used to reduce and mitigate the threat of participants not reflecting the target population needs to be detailed, since, following a convenience sampling strategy, links to the survey were spread via multiple networks and mailing lists (Section 3.2.2). The terminology used in the survey required specific knowledge to answer the questionnaire, and the consistently meaningful free-text answers, which have been analyzed qualitatively in Klünder et al,16 indicate that this threat if present had an insignificant impact. On this regard, it is important to highlight that no definition was provided for any given item, be this a method, a practice, or a technical concept in general. While this was done to avoid any bias that might have been introduced from exposing participants to our knowledge, informing them on the concepts of interest, or leading them towards specific attitudes when answering the survey questions, the lack of definitions also provided a barrier to engage with the survey that reduced the risk of receiving answers from participants outside the target population.

**5.3.2 Internal validity**

Prior to the analysis, we cleaned the data (Section 3.3.1), which could introduce a threat to internal validity as errors might have been introduced. Also, in the data analysis, we did not exclude data points per sé, but performed the analyses with varying $n$'s. To mitigate the risks, all steps have been performed by at least two researchers and have been checked by other researchers not involved in the actual analysis activities. Due to these review processes, we have confidence that the method is reliable and reproducible.

**5.3.3 Conclusion validity**

The interpretation of the statistical tests is based on a significance level of $p \leq 0.05$, and we found no evidence that allows us to reject our null hypotheses (Table 1). Furthermore, for analyzing sets of methods and practices, we used a 35% and an 85% threshold (Section 3.3.3). Changing these thresholds would influence the results by enlarging the sets of methods and practices. Also, the limited set of options for the multiple- choice questions could influence the findings. The choice of the thresholds can of course be discussed. However, we contend that the effects observed lay the foundation for future research, which is necessary to study the effects in more detail.

**5.3.4 | External validity**

Although our analysis is based on a large dataset (Section 3.3.1), we cannot claim full generalizability. Yet, we reached a broad coverage of domains and participant roles as well as an even distribution of company sizes (Section 4.1). This allows for making observations that are independent of these factors. For other factors, further research is necessary. Nevertheless, the generalization of a single study to all cases of software development is a threat. Moreover, concerning the generalizability of results across countries it would have been necessary to have more data points from Africa, Asia, and North America (Figure 2). Having few data points from countries in these regions threatens the global generalizability of our results. However, the data points that we have, for example, from Uganda, indicate that our results might be to some degree valid for these regions as well. Future studies are needed to confirm this.

---

[2] Taken from an interview by informIT, March 23, 2001: https://www.informit.com/articles/article.aspx?p=20972, last access: February 6, 2019.

## 6. CONCLUSION

In this paper, using a large-scale international online survey, we studied the use of *hybrid development methods* in practice. An analysis of 1467 data points revealed that using different frameworks, methods, and practices in combination as hybrid methods is the norm across companies of all sizes and industry sectors. We identified eight base methods and few practices only that find agreement among study participants. For the study participants that explicitly stated to use processes in combination, we could identify 27 base methods and method combinations that, together with three practices forming three pairs, build the basis to devise hybrid methods. We also found that the sets of practices have limited dependencies to the methods. We therefore argue that practices are the building blocks for devising hybrid methods.

In terms of future research, we plan to build on our observations and findings showing that practices are the essential unit of analysis when looking at software development activities within an organization. We note the core set of practices along with the complementary sets of practices identified in Section 4.5 are common to all development methodologies. Because they are so widely deployed, we observe that development organizations see these practices as essential activities enabling them to deliver good software to their customers. We believe that the idea of having a set of common practices that are essential to sound software development has been the motivation behind maturity model frame- works like the CMMI, ISO/IEC 15504, and others. For our future work, we would like to conduct further analysis using the HELENA dataset to explore what having a core set of practices means regarding how industry views the value of maturity model frameworks and specific key process areas within those frameworks. Finally, a future direction of investigation will involve the slicing of the dataset into all its variables to identify whether statistically influencing parameters in the context described by participants can be used to identify recurring inclinations in the choices of hybrid methods. Therefore, answering questions similar to the following: *Does the application domain influence the choice of hybrid methods and how?* or *Does the criticality of the software systems influence the choice of hybrid methods and how?*

**ACKNOWLEDGEMENTS**

We thank all the study participants and the researchers involved in the HELENA project for their great effort in collecting data. *Dietmar Pfahl* was supported by the institutional research grant PRG887 of the Estonian Research Council as well as the Estonian IT Center of Excellence (EXCITE) TK148.